\documentclass[preprint, aps, pra]{revtex4-2}
\usepackage{amsmath}
\usepackage{amssymb}
\usepackage{hyperref}
\usepackage{amsmath}
\usepackage{graphicx}
\usepackage{xcolor}
\usepackage{floatpag}
\usepackage{geometry}
\begin{document}

\title{Antiskyrmions and their electrical footprint in crystalline mesoscale structures of \texorpdfstring{Mn$_{1.4}$PtSn}{Mn1.4PtSn}} 

\author{Moritz Winter$^{1,2,3}$, Francisco J. T. Goncalves$^{4}$, Ivan Soldatov$^{5}$, Yangkun He$^{2}$, Bel\'en E. Z\'u\~niga C\'espedes$^{2,6}$, Peter Milde$^{6}$, Kilian Lenz$^{4}$, Sandra Hamann$^{1}$, Marc Uhlarz$^{1}$, Praveen Vir$^{2}$, Markus K\"onig$^{2}$, Philip J. W. Moll$^{2,7}$, Richard Schlitz$^{8}$, Sebastian T. B. Goennenwein$^{8}$, Lukas M. Eng$^{6,9}$, Rudolf Sch\"afer$^{5,8}$, Joachim Wosnitza$^{1,8,9}$, Claudia Felser$^{2,9}$, Jacob Gayles$^{2,10,\ast}$, and Toni Helm$^{1,2,\ast\ast}$\\
{$^{1}$Dresden High Magnetic Field Laboratory (HLD-EMFL), Helmholtz-Zentrum Dresden-Rossendorf, 01328 Dresden, Germany.}\\
{$^{2}$Max Planck Institute for Chemical Physics of Solids, 01187 Dresden, Germany.}\\
{$^{3}$Dresden Center for Nanoanalysis, cfaed, Technical University Dresden, 01096 Dresden, Germany.}\\
{$^{4}$Institute of Ion Beam Physics and Materials Research, Helmholtz-Zentrum Dresden-Rossendorf, 01328 Dresden, Germany.}\\
{$^{5}$Leibniz Institute for Solid State and Materials Research Dresden, 01069 Dresden, Germany.}\\
{$^{6}$Institute of Applied Physics, Technical University Dresden, 01062 Dresden, Germany.}\\
{$^{7}$\'Ecole Polytechnique F\'ed\'erale de Lausanne, 1015 Lausanne, Switzerland.}\\
{$^{8}$Institute for Solid State and Materials Physics, Technical University Dresden, 01062 Dresden, Germany.}\\
{$^{9}$W\"urzburg-Dresden Cluster of Excellence ct.qmat, 01062 Dresden, Germany.}\\
{$^{10}$Department of Physics, University of South Florida, Tampa, Florida 33620, USA.}\\
{$^\ast$E-mail: gayles@usf.edu}\\
{$^{\ast\ast}$E-mail: t.helm@hzdr.de}\\
}
%\date{\today}

\begin{abstract}
Skyrmionic materials hold the potential for future information technologies, such as racetrack memories. 
Key to that advancement are systems that exhibit high tunability and scalability, with stored information being easy to read and write by means of all-electrical techniques. 
Topological magnetic excitations such as skyrmions and antiskyrmions, give rise to a characteristic topological Hall effect. 
However, the electrical detection of antiskyrmions, in both thin films and bulk samples has been challenging to date. 
Here, we apply magneto-optical microscopy combined with electrical transport to explore the antiskyrmion phase as it emerges in crystalline mesoscale structures of the Heusler magnet Mn$_{1.4}$PtSn. 
We reveal the Hall signature of antiskyrmions in line with our theoretical model, comprising anomalous and topological components. 
We examine its dependence on the vertical device thickness, field orientation, and temperature. 
Our atomistic simulations and experimental anisotropy studies demonstrate the link between antiskyrmions and a complex magnetism that consists of competing ferromagnetic, antiferromagnetic, and chiral exchange interactions, not captured by micromagnetic simulations.
\end{abstract}
\maketitle 

%\section*{Main Text}

\section*{Introduction}
The field of skyrmionics comprises new phases of magnetic systems, where the individual spins align as whirls \cite{Bogdanov1989,Bogdanov1994,Roessler2006,Back2020}, that hold the potential to advance the understanding of topology in condensed-matter physics. Fundamentally, magnetic textures, such as skyrmions \cite{Roessler2006, Muehlbauer2009}, antiskyrmions \cite{Nayak2017, Koshibae2016} (ASKs) and related skyrmionic systems \cite{Rybakov2015,EverschorSitte2018,Goebel2021}, are distinguished by the relative rotation of the individual magnetic moments with respect to one another in the presence of a magnetically ordered background. 
This topologically protected state can be used to robustly carry and store high-density information at fast speed with low power consumption, as proposed in racetrack memory spintronic devices \cite{Nagaosa2013,Parkin2015}.
Solitary skyrmion states as well as crystal lattices of skyrmion systems have been reported for many magnetic crystals that break inversion symmetry such as bulk chiral magnetic compounds \cite{Muehlbauer2009,Neubauer2009}, multilayer heterostructures \cite{Yang2020,Moreau2016}, thin films \cite{Yu2011}, oxides \cite{Seki2012}, and, more recently, tetragonal Heusler systems \cite{Nayak2017,Jena2020}. 
While experimental observations from, e.g., Lorentz transmission electron microscopy \cite{Nayak2017,Shimojima2021,saha2019intrinsic}, neutron scattering \cite{Muehlbauer2009}, and magnetic force microscopy \cite{Milde2013,Zuniga2021} have shown great advancement over the last decades, detection by electrical means is required for the realization of energy-efficient spintronic devices \cite{Nagaosa2013,Parkin2015}, probabilistic \cite{Zazvorka2019} and neuromorphic computing \cite{Song2020}. 
Furthermore, a pivotal point for applications is the ability to control and scale skyrmionics \cite{Yu2020,Han2020}, or even to transform one species of magnetic texture into another in multiskyrmion systems \cite{Jena2020,Ma2020,Karube2021}. 

The tetragonal Heusler compound Mn$_{1.4}$PtSn \cite{Vir2019b} has recently gained a lot of interest as it hosts distinct topological states. 
Most excitingly, an ASK lattice can be established at room temperature. 
Besides the high tunability of the Heusler compounds \cite{Peng2020,Ma2020,Zuniga2021}, the competition of magnetic exchange interactions and anisotropy lead to an enhanced temperature range over which ASKs can be stabilized in comparison to skyrmion bulk systems. 
Bulk Mn$_{1.4}$PtSn displays a complex correlation of exchange between two Mn sublattices that leads to a spin reorientation from a collinear to a noncollinear arrangement at temperatures $T_\mathrm{SR}\approx 170\,$K and below, which contributes to the anisotropy at low temperatures. 
At temperatures above the spin reorientation, sophisticated magnetic textures can be realized thanks to the combination of in-plane (perpendicular to the $c$ axis) and anisotropic Dzyaloshinskii-Moriya interaction (DMI), crystal anisotropy along the tetragonal axis, and dipole-dipole interaction due to large moments ($>4\mu_\mathrm{B}$). 
Above $T_\mathrm{SR}$, in the absence of an external magnetic field, the compound prefers a spin-spiral state that is directly connected to the ratio of the exchange interaction and DMI. 
For a finite range of external magnetic fields, a long-range ($>100\,$nm) hexagonal ASK phase as well as a short-range noncoplanar state ($<1\,$nm) were found \cite{Nayak2017}.

These ASKs exhibit an extraordinary stability to sample-thickness variations to at least a few hundred nanometers confirmed by Lorentz transmission electron microscopy \cite{saha2019intrinsic} and x-ray studies \cite{Seki2022direct}. 
The latter confirmed bulk skyrmions in few microns thick samples. 
The diameter of the ASKs scales linearly with thickness and was found to grow to multiple hundreds of nanometers in micron-sized samples \cite{Ma2020,Zuniga2021}. For very thin platelets, the phase diagram becomes more complex. 
Multiple skyrmion-like textures that are different above and below $T_\mathrm{SR}$ can be stabilized by varying the field orientation \cite{Peng2020,Jena2020}. 
Surprisingly, neutron-scattering experiments on bulk Mn$_{1.4}$PtSn did not resolve any ASKs \cite{Sukhanov2020}. 
Even more intriguing, a huge additional component in the Hall effect, associated with a topological origin, emerges in the low-temperature region of the phase diagram, just below $T_\mathrm{SR}$, exceeding expectations for skyrmions \cite{Vir2019a}. 
This is likely related to noncollinear magnetism. 
Therefore, it is still an open question if and how ASKs contribute to the Hall effect and, in particular, how strong that effect is.

The topological nature of skyrmions and ASKs should be observable in terms of a unique response to an external electric field, i.e., a topological Hall effect (THE) in electrical transport. 
The THE belongs to the family of Berry-curvature Hall effects \cite{Lux2018,Lux2020}, where emergent fields due to a real-space variation of the local magnetism, describable by the Berry curvature \cite{Berry1984}, are the origin of a transverse voltage. 
Closely related to the THE is the anomalous Hall effect (AHE), due to the interaction of localized and conduction electrons \cite{Nagaosa2010}. 
In a finite magnetic field, the Hall response is a voltage transverse to the applied current. 
This is sketched in Fig.~\ref{fig1}a. 
Phenomenologically, the total Hall effect can be described by a superposition of three contributions \cite{Nagaosa2010}:
\begin{equation}
 \rho_{xy}=\rho^\mathrm{OHE}_{xy}+\rho^\mathrm{AHE}_{xy}+\rho^\mathrm{THE}_{xy}.
\end{equation}
The ordinary Hall effect (OHE) that scales with the external magnetic field, $\mathbf{H}$; The AHE, which scales with the magnetization $\mathbf{M}$, describable in terms of an intrinsic momentum-space Berry curvature and extrinsic scattering mechanisms, such as skew and side-jump scattering \cite{Nagaosa2010, Bruno2004}; The THE, which can be described in terms of the  chiral product:
\begin{equation}
  \mathbf{\chi}_{ij}=\mathbf{S}_i\cdot(\mathbf{S}_j\times \mathbf{S}_k),
\end{equation}
where $\mathbf{S}_i$ is the local spin orientation. 
The latter may be utilized to identify the sign of the topological charge, being opposite for skyrmions and ASKs, respectively.

In this work, we successfully combined \textit{in-situ} transport with magneto-optical microscopy applied to mesoscale structures fabricated from single-crystals by focused-ion-beam assisted patterning. 
This approach enables us to directly examine the Hall signature of the ASK phase in Hall-bar devices with thicknesses ranging from $500\,$nm to $10\,\mu$m, observable from $T_\mathrm{SR}$ to room temperature and above. 
The antiskyrmionic Hall component $\rho^\mathrm{ASK}_{xy}= \rho^\mathrm{AHE}_{xy}+\rho^\mathrm{THE}_{xy}$ is comprised of an anomalous and a topological component. We study its temperature, field-orientation, and device-thickness dependence. 
As the expected THE is rather small, we conclude that the major contribution originates from the AHE of the ASKs. 
In contrast to the general approach of micromagnetic simulations, we apply atomistic spin-dynamic calculations that capture the underlying complex magnetism of Mn$_{1.4}$PtSn, inevitably linked to the emergence of ASKs at high temperatures. 
We show experimentally how the periodicity of magnetic textures and the $\rho^\mathrm{ASK}_{xy}$ of ASKs scales with thickness. 
Our ferromagnetic resonance (FMR) studies reveal the distinct magnetic anisotropies above and below $T_\mathrm{SR}$ that lay the foundation for the emergence of multiple skyrmionic textures reported previously.

\section*{RESULTS AND DISCUSSION}
\subsection*{Room-temperature Hall signature of the antiskyrmion phase}
\begin{figure}[tb]
	\centering
	\includegraphics[width=0.85\linewidth]{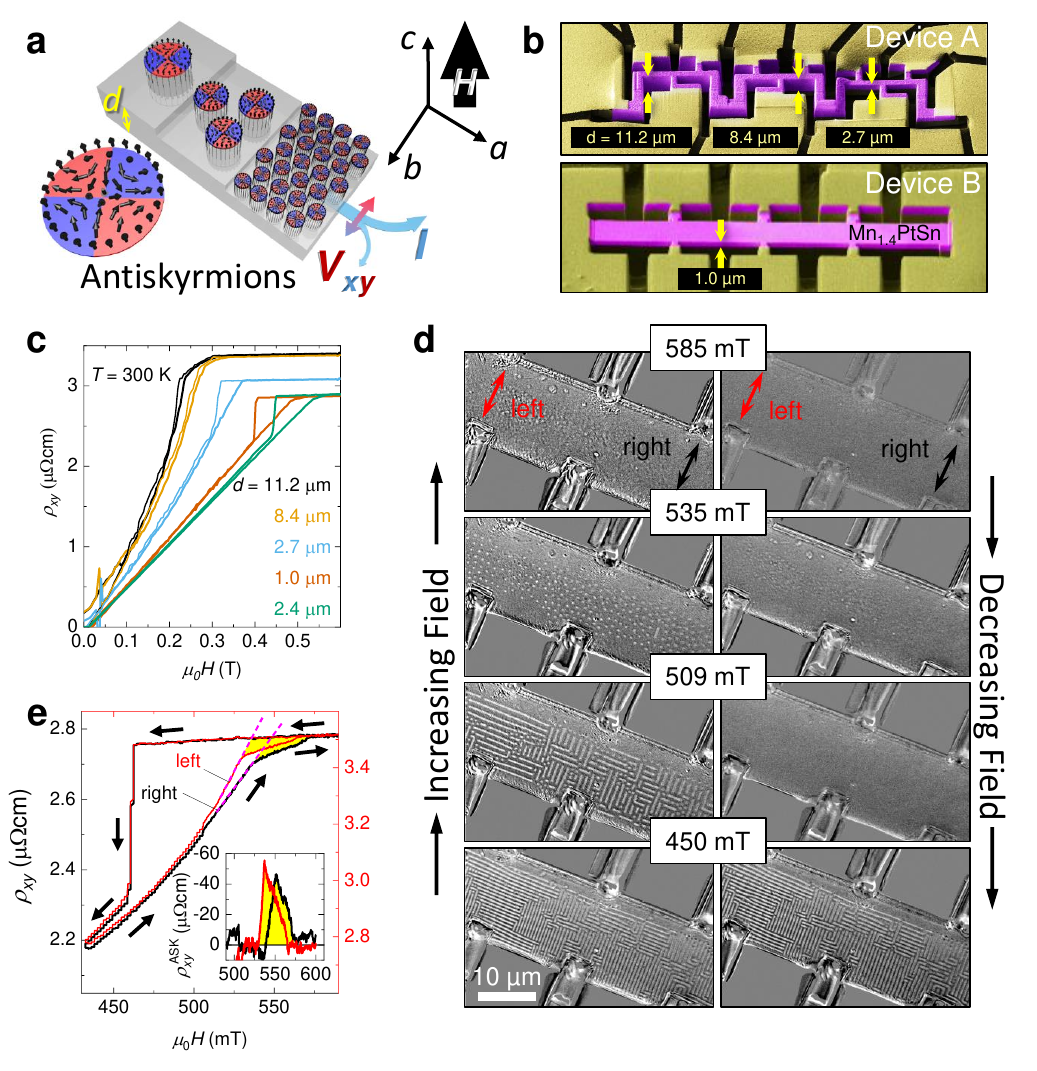}
	\caption{{\bf Transport signature of ASKs in micro-fabricated Mn$_{1.4}$PtSn devices.} (a) Schematic visualization of thickness-dependent ASKs induced by magnetic field, $H \parallel c$ axis, in a step-like sample with applied current along the $a$ axis. (b) False-color SEM images of FIB devices A and B with meander-shape (three thicknesses: $d=11.2$, 8.4, and $2.7\,\mu$m) and Hall-bar geometry ($d=1.0\,\mu$m), respectively, contacted via sputter-deposited gold contacts. (Further details of all investigated devices, dimensions, and fabrication are presented in Supplementary Notes\ 1-3). (c) Hall-resistivity loop of samples A, B, and C with different thicknesses, $d$, along the $c$ direction with $H \parallel c$ at $T = 300\,$K. (d) Greyscale MOKE images of Hall-bar transport device F ($d=0.8\,\mu$m) at four different fields for up and down sweep, respectively. (e) \textit{In-situ} Hall loop for the left and right contacts of device F marked by red and black color, respectively.
	}
	\label{fig1}
\end{figure}

In Fig.~\ref{fig1}a, we show a schematic of a step-like device, where an ASK lattice is formed in an external magnetic field applied perpendicular to the $ab$ plane of the step device (for further details see Supplementary Notes\ 1-3).
ASKs possess an anisotropic winding with a topological charge of opposite sign as compared to isotropic Bloch and N\'eel skyrmion systems \cite{Koshibae2016}.
As the thickness, $d$, of a device made from Mn$_{1.4}$PtSn is reduced the periodicity of the ASK lattice decreases and, hence, the ASK size, too.
This in turn increases the density and should, therefore, directly affect the magnitude of the topological charge that contributes to the Hall voltage, $V_{xy}$, transverse to the electric current, i.e., the THE. 
Furthermore, the local net magnetization of an individual ASK is roughly zero. 
Consequently, the volume magnetization should exhibit a reduction as ASKs emerge. 
This would affect the anomalous Hall effect, which is proportional to the magnetization~\cite{zeissler2018discrete,maccariello2018electrical}. 
However, direct measurements of the magnetization for sub-micron-thin, small-in-volume samples of Mn$_{1.4}$PtSn remains a technical challenge.

Figure~\ref{fig1}b displays false-color scanning electron microscope images of two different devices fabricated by focussed-ion-beam (FIB) assisted patterning, namely, a step-like device with thicknesses $d= 11.2$, $8.4$, and $2.7\,\mu$m and a standalone device with $d=1.0\,\mu$m. 
In Fig.~\ref{fig1}c, we display the total measured Hall effect for all thicknesses from the demagnetized state ($\mu_0 H=0$) up to a nearly saturated state ($H_\mathrm{S}$). 
For $d < 8.4\,\mu$m, a noticeable hysteresis appears in the Hall effect before saturation, which details a noncoplanar magnetic and ASK spin texture, when the field is applied parallel to the $c$ axis. 
Device F with $d=0.8\,\mu$m was set up as a Hall bar to simultaneously measure the Hall voltage while observing the magnetic state \textit{in-situ} via polar magneto-optical Kerr effect (MOKE) microscopy \cite{Soldatov2018}. 
Selected MOKE images of the device are presented in Fig.~\ref{fig1}d. The Hall data in Fig.~\ref{fig1}e was simultaneously measured with polar MOKE microscopy, shown in Fig.~\ref{fig1}d as a function of external field. 
Fig.~\ref{fig1}e displays the Hall-resistivity hysteresis loops recorded for the left (red) and the right (black) contacts (shown in Fig.~\ref{fig2}f) of device F with the sweep direction indicated by black arrows. 
The down sweep stays in the saturated state down to approximately 460\,mT due to the anisotropy in the magnetic exchange interactions and the critical aspect ratio of the sample that adds an additional shape-anisotropy contribution \cite{Zuniga2021}. 
However, on the up sweep there are two clearly distinguishable slopes: The first originates from distinct conical magnetic spin spirals formed by the two non-equivalent magnetic Mn sublattices \cite{Vir2019a}. 
This hysteresis is also discernible in the magnetization, as we show in Supplementary Note\ 4. 
The second, highlighted by the yellow-shaded area, shows a slope change and coincides with the formation of the ASK lattice. 
The MOKE images of Fig.~\ref{fig1}d, recorded simultaneously with the Hall resistivity, show a saturated magnetized state down to above 450\,mT, where band domains associated with magnetic spin-spirals suddenly emerge. 
When increasing field from zero, the conical spin-spiral state starts to slowly disintegrate into individual ASKs and ASK strings. 
Then suddenly at 535\,mT, an ASK lattice emerges, exactly where the shoulder in the Hall resistivity develops. 
A more refined video sequence that shows a full field scan between saturation and zero field can be found in Supplementary Movie 1. 
In Supplementary Note\ 5, we present an additional sequence of MOKE images recorded for device C with $d=2.4\,\mu$m before it was structured into the Hall-bar geometry, where we observed much larger band domains and consequently a less dense ASK lattice.

\subsection*{The magnetic structure of \texorpdfstring{Mn$_{1.4}$PtSn}{MnPt1.4Sn} and the expected transport signature in the ASK phase}
\begin{figure}[tb]
	\centering
	\includegraphics[width=0.9\linewidth]{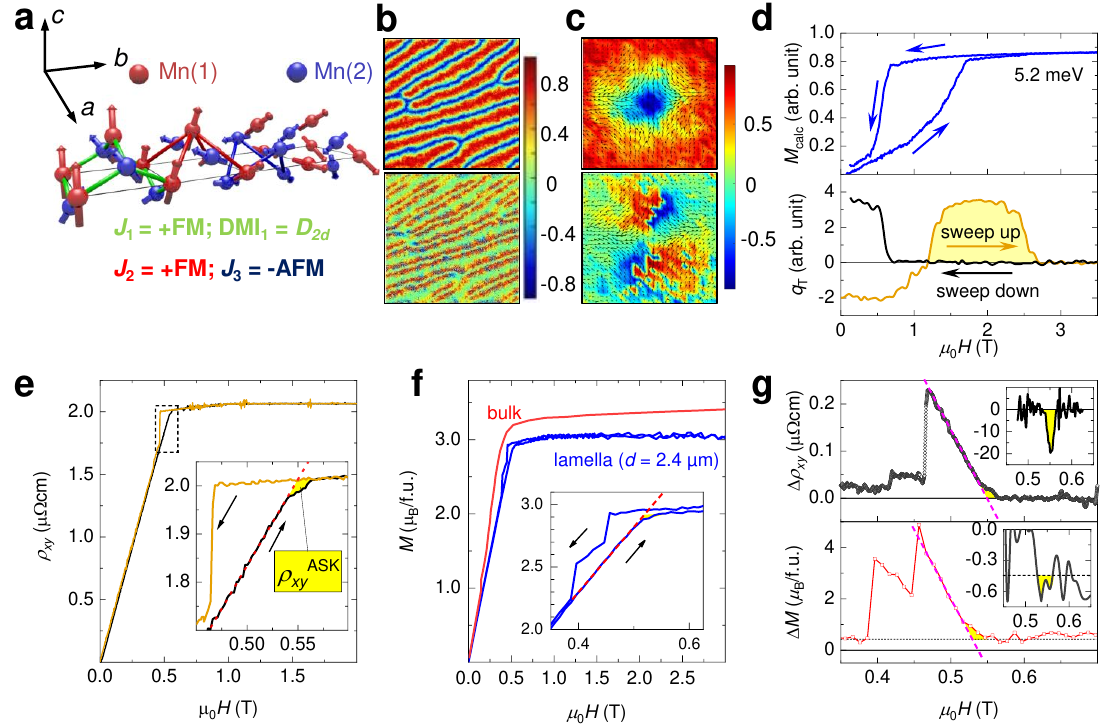}
	\caption{{\bf Magnetic structure of Mn$_{1.4}$PtSn.} (a) Sketch of the relevant interaction and coupling constants linking the two Mn sublattices (red and blue). (b), (c) Atomistic spin-dynamic simulations for $H \parallel c$: Color maps for an array of $(372 \times 372)$ unit cells showing the perpendicular $z$ component in the spin-spiral phase (left) and ASK phase (right), respectively. Black arrows represent the in-plane component. The color scale is given in radiant, where 1/-1 corresponds to a spin alignment parallel/antiparallel to the normal. Lower panels show the in-plane spin-winding component. (d) Upper panel: Calculated magnetization loop for fixed temperature given in meV; Lower panel: Calculated topological charge for opposite field-sweep directions (orange and black line). (e) Full up and down Hall-resistivity traces for device C ($d=2.4\,\mu$m) recorded at 300 K. Inset: Reduced-area plot. Arrows indicate the field-sweep direction. Red dashed line is a linear fit. (f) Magnetization for a bulk sample and device C, recorded before it was structured into a Hall bar by FIB. Inset: Reduced-area plot. (g) Differences between up and down sweeps for the Hall and magnetization data shown in (e) and (f), respectively. Insets: Background-subtracted component related to the field region where ASKs are detectable.
	}
	\label{fig2}
\end{figure}

In Fig.~\ref{fig2}a, we sketch the spin configuration and the respective interactions present in Mn$_{1.4}$PtSn.
The combination of DMI, induced by the D2d symmetry, FM and AFM interaction, is an ideal foundation for diverse magnetic textures \cite{Vir2019a}. 
We find that the physics of topological textures in Mn$_{1.4}$PtSn can be accurately modeled purely from the exchange interactions and magnetocrystalline anisotropy in three atomic layers of the two magnetic sublattices \cite{Goebel2021,Akosa2018, Caretta2018}. 
More significantly, the ratios of the exchange constants determine the underlying physics; therefore, we tuned the classical parameters in atomistic spin-dynamic simulations to clearly show the effects seen in the experiments. 
There are three sets of exchange interactions and DMI crucial for the stabilization of in-plane ASKs in Mn$_{1.4}$PtSn. 
The first interaction, $J_1$ between the Mn(1) and Mn(2) sublattice, is ferromagnetic. It reaches along the shortest magnetic distance, where the DMI is strongest. 
The second ferromagnetic interaction, $J_2$ between magnetic sublattices of the same type and perpendicular to the $c$ axis, is the main interaction responsible for the Curie temperature. 
The interaction $J_3$ for the Mn sublattice Mn(2) at the Wyckoff position $4d$ is of AFM nature, and assists in the formation of the noncoplanar structure below $T_\mathrm{SR}$. 
These AFM and FM interactions compete to form double-chiral spin spirals in order to minimize the energy. The chirality degeneracy is lifted by the DMI.

The calculated out-of-plane (in-plane) component of the spin-spiral and ASK state is shown in the upper (lower) panel in Fig.~\ref{fig2}b and \ref{fig2}c, respectively. 
The yielded magnetization hysteresis (Fig.~\ref{fig2}d) is in line with our experimental observations of a finite hysteresis away from zero field for a transition from the spin-spiral ground state into a stable ASK phase (for further details see Supplementary Note\ 6). 
The calculated magnetization exhibits a weak region, that is a shallow slope change right before the saturation field, consistent with our THE observations (see Fig.~\ref{fig2}d). 
The chiral product, $\chi_{ij}$ leads to a topological winding number, $w=1/4\pi\int{\mathbf{\hat{M}}\cdot (\delta_x\mathbf{\hat{M}} \times \delta_y \mathbf{\hat{M}})}dxdy$, where $\delta_i=\delta/\delta_i$, and $\mathbf{\hat{M}}(r,t)=\mathbf{M}(r,t)/|\mathbf{M}|$ is the direction of the magnetization at each spatial position. 
The presence of $w$ causes an emergent magnetic field, $H^e=\frac{\hbar}{2}\epsilon_{zxy} \mathbf{\hat{M}}\cdot (\delta_x \mathbf{\hat{M}} \times \delta_y \mathbf{\hat{M}})$, originating from a real-space Berry curvature \cite{Lux2018}. 
The so-called topological charge, $q_\mathrm{T}$, is negative for skyrmions and positive for ASKs. 
Thereby, the THE is a direct link to the topology of the magnetic texture:
\begin{equation}
\rho_{xy}^\mathrm{THE}=R_{xy}^\mathrm{THE}H^e,
\end{equation}
where the Hall coefficient, $R_{xy}^\mathrm{THE}$, is tied to the complex multiorbital electronic structure \cite{Spencer2018}. 
The composite nature of $R_{xy}^\mathrm{THE}$ and $H^e$ complicates the differentiation of skyrmionic configurations with distinct winding numbers across multiple materials. 
However, in Mn$_{1.4}$PtSn distinct magnetic textures are induced by the external field.

In the lower panel of Fig.~\ref{fig2}d, we plot $q_\mathrm{T}$ against the external field. 
In the field-up sweep, $q_\mathrm{T}$ in the ASK phase exhibits an opposite sign as compared to the net charge in the low-field spin-spiral phase and extends to higher fields as compared to the reoccurring $q_\mathrm{T}$ in the down sweep (orange curve). 
The down sweep (black curve) remains flat and only exhibits a positive contribution at lower fields, originating from the reestablishing spin-spiral phase. 
Hence, the subtraction of field-up and -down sweep mostly cancels out the component from the ordered phase and should, therefore, yield the Hall effect due to ASKs directly. 
In Mn$_{1.4}$PtSn, ASK sizes can reach a few hundred nanometers in diameter. 
For these huge objects the expected THE is rather small \cite{zeissler2018discrete}. 
Figures~\ref{fig2}e, f, and g provide a comparison of the signatures observed in the Hall resistivity as well as in the magnetization for a $2.4\,\mu$m thick lamella sample. 
The hysteresis shows up for both quantities and a linear fit of the slope in the differences $\Delta\rho_{xy}$ and $\Delta M$ just below the saturation field reveals a shoulder-like feature being discernible in both quantities. 
The insets of Fig.~\ref{fig2}g present the contribution we associate with the presence of ASKs. 
Apparently, the shoulder can be addressed to an anomalous Hall component due to the emergence of the ASK textures. 
The magnetic moment scales with the volume of the material and the observed feature in the magnetization is hardly distinguishable from the noise background. 
We also know that ASKs only show up for micron-thick devices. 
Therefore, it is hard to trace the ASK feature for smaller thicknesses in the magnetization. 
Here, the Hall effect appears to be the ideal tool. 
In Supplementary Note\ 7 we provide subtraction results from our experimental transport data at various temperatures for device B with $d=1\,\mu$m. 
A clear feature can easily be traced upon varying the device thickness, temperature, and field orientation (see the following sections).

\subsection*{Thickness- and temperature-dependent Hall effect in the ASK phase}
 
\begin{figure}[tb]
	\centering
	\includegraphics[width=0.7\linewidth]{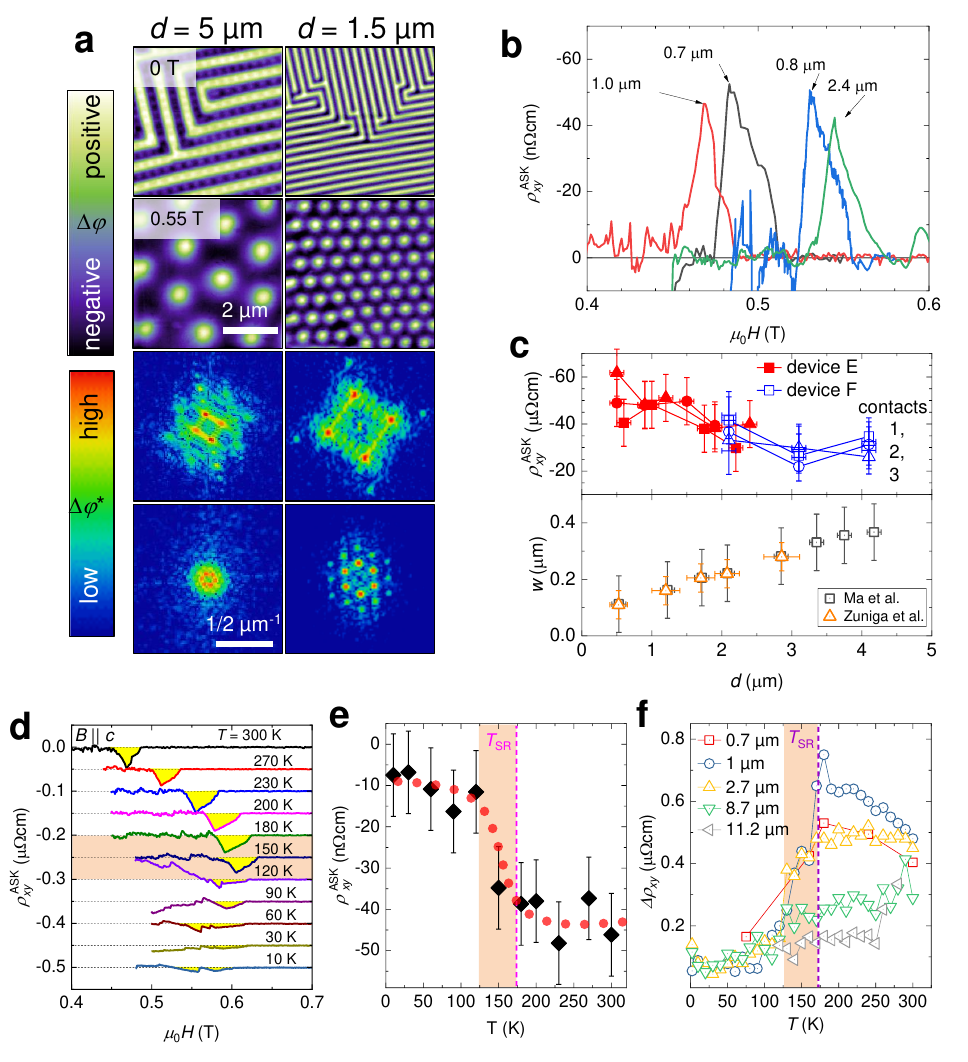}
	\caption{{\bf Thickness and temperature dependence of the magnetic textures and Hall component associated with ASKs.} (a) MFM images (local magnetic $z$ component represented by $\Delta\phi$ showing chiral domains and an ASK lattice for two samples prepared by FIB patterning with thicknesses of 5 and $1.5\,\mu$m. The data were recorded at room temperature in zero-field and at 550\,mT applied parallel to the $c$ axis, i.e., parallel to the line of sight. The lower panels show the respective FFTs (in arb. unit with logarithmic color scale). (b) ASK Hall component, $\rho_{xy}^{\mathrm ASK}$, plotted against magnetic field for devices B, C, E, and G with thicknesses of 1, 2.4, 0.7, and $0.8\,\mu$m, respectively. (c) Thickness dependence of $\rho_{xy}^{\mathrm ASK}$ compared to the domain periodicity, $w$, determined by recent MFM measurements \cite{Ma2020,Zuniga2021}. The error bars represent the standard deviation of the values for variations in the details of the background subtraction. (d) $\rho_{xy}^{\mathrm ASK}$ for temperatures between 10 and 300\,K for device B with $d = 1\,\mu$m. (e) Maximum amplitude of $\rho_{xy}^{\mathrm ASK}$ in device B ($d=1\,\mu$m) plotted against temperature. Red dots are a guide to the eye. The error bars are defined as in (c).  (f) Temperature dependence of the maximum difference in the Hall-resistivity loop, $\Delta\rho_{xy}$ between up and down sweeps for devices A, B, and E.
	}
	\label{fig3}
\end{figure}

 While MOKE microscopy in combination with electrical transport allows for simultaneous measurements of the magnetic texture and the Hall effect in Mn$_{1.4}$PtSn, it is limited in spatial resolution by the wavelength of the optical light used \cite{Soldatov2018}. 
 We, therefore, use magnetic force microscopy (MFM) to resolve the ASK lattice induced in samples of different thickness $d$. 
 Figure~\ref{fig3}a shows MFM images at zero and 0.55\,T for a $5\,\mu$m and $1.5\,\mu$m thick sample. 
 The spin-spiral domain bands and the ASK lattice are discernible with varying periodicity, depending on $d$ (the lower $d$ the smaller the magnetic periodicity). 
 The fast Fourier transforms (FFTs), shown in the lower panels in Fig.~\ref{fig3}a, highlight the change in topology as the system transforms from the spin-spiral into the hexagonal ASK phase. 
 The size and periodicity of the ASKs are affected by the thickness as has been shown recently–the latter varies almost linearly with thickness \cite{Zuniga2021,Sukhanov2020}.
 
 This suggests that a THE induced by ASKs is expected to grow quadratically with decreasing $d$, as it is directly related to the ASK density, which is in turn proportional to the area occupied by each ASK. 
 In Fig.~\ref{fig3}b, we show $\rho^\mathrm{ASK}_{xy}$ at room temperature determined for various devices with different thicknesses between 2.4 and $0.8\,\mu$m. 
 The overall magnitude varies between 40 and 50\,n$\Omega$cm. We also studied the thickness dependence in more depth on transport devices E and G, presented in Fig.~\ref{fig3}c. 
 In these cases, we lowered the thickness by low-energy (5\,kV) Ar-ion etching in steps and measured the room-temperature Hall effect after each thinning step. 
 The overall trend of an increased response for smaller thickness is apparent. 
 However, this may also originate from an increasing number of ASKs that can emerge in between the electrical Hall contacts due to their shrinking diameters upon device-thickness reduction. 
 As was shown already for N\'eel skyrmions in Co nanolayers, the AHE of single skyrmions may be much stronger than the THE component\ \cite{zeissler2018discrete}. 
 Therefore, at this point, we can only provide an upper estimate of the THE: Namely, it must be smaller than a fraction of the magnitude of the detected $\rho^\mathrm{ASK}_{xy}$. 
 In Supplementary Note 7 we provide a theoretical estimate following the previous approaches\ \cite{Muehlbauer2009,Spencer2018} for a known ASK density. 
 For our microscale devices the ASK size is approximately 100\,nm, which would lead to a theoretically expected THE component of $| \rho^\mathrm{THE}| \sim 50\,\mathrm{n}\Omega$cm.
 
 Moreover, we find that $\rho^\mathrm{ASK}_{xy}$ remains as pronounced as at room temperature all the way down to temperatures close to $T_\mathrm{SR}$ (see Fig.~\ref{fig3}d and Supplementary Note\ 7). 
 Below $T_\mathrm{SR}$, it is strongly suppressed until it vanishes and additional step-like jumps in the Hall resistivity occur. 
 The saturation field marks the upper boundary of the ASK field region (yellow). 
 With decreasing temperature, it shifts towards higher values until saturating near $T_\mathrm{SR}$, where noncoplanar magnetism starts to take over. 
 Its overall amplitude starts to subside around $T_\mathrm{SR}$ as can be seen from Fig.~\ref{fig3}e. 
 Below 150\,K, we cannot unambiguously link the observed deviations to skyrmions. 
 To visualize the crossover around $T_\mathrm{SR}$ we show the maximum  $\Delta\rho_{xy}$ plotted against the temperature in Fig.~\ref{fig3}f. 
 This demonstrates the gradual change of the overall behavior of the observed hysteresis near the spin-reorientation transition region (shaded region) associated with the onset of noncoplanar order and a reorientation of the magnetic easy axis. 
 Further microscopic and spectroscopic studies are highly desirable in order to understand the details of the temperature dependence.

\subsection*{Emergence of a new state below \texorpdfstring{$T_\mathrm{SR}$}{TSR}}
Below $T_\mathrm{SR}$, the overall Hall response decreases more rapidly and the hysteresis in field narrows (see Fig.~\ref{fig4}a). 
We present a detailed angular study on device B in Supplementary Note\ 8. 
As we tilt the magnetic field away from $H \parallel c$, the hysteresis subsides (it is absent in the high-temperature curves shown in Fig.~\ref{fig4}b). 
For temperatures near and below $T_\mathrm{SR}$ and the field being oriented away from the $c$ direction, we observe a prominent hump-like enhancement of $\rho_{xy}(H)$ close to the saturation field (see for example the maximum in $\rho_{xy}(H)$ at 180\,K and $H \parallel b$ in Fig.~\ref{fig4}b). 
This feature was already explored for bulk samples \cite{Vir2019a, Vir2019b}. 
Its origin is a THE linked to the establishment of a noncoplanar spin structure with a strong real-space Berry curvature. 
Excitingly, for fields along the $b$ axis, $\rho_{xy}(H)$ starts acquiring a negative high-field slope and even changes sign at low temperatures (Fig.~\ref{fig4}b). 
At 2\,K, we observe a surprisingly large transverse transport signal even without any out-of-plane field component. 
For angles within only a few degrees off the in-plane orientation (i.e., $\theta=90^\circ$) it is accompanied by a new hysteresis that exhibits opposite sign as compared to the one for $H \parallel c$. 
As we fine-tune the angular step width, we observe a hysteresis that extends to fields even larger than 4\,T, being extremely sensitive to minute changes in $\theta$ (see Supplementary Note\ 8). 
Moreover, step-like changes in the hysteretic part of $\rho_{xy}(H)$ occur (see for example the 150\,K curve in Fig.~\ref{fig4}b). 
Hence, our data indicates an intriguing magnetic behavior for low temperature and field aligned within the ab planes, likely originating from the noncoplanar magnetism with a much stronger magnetocrystalline anisotropy for low temperatures. 
As we show in Fig.~\ref{fig4}d and Supplementary Note\ 8, both the AHE and the OHE deviate from the conventional $\cos{\theta}$ dependence, observable at room temperature, as temperature is tuned to $T_\mathrm{SR}$ and below (see red dashed fits). 
This further indicates an enhanced saturation field for the in-plane field orientation. 
These temperature- and angle-dependent changes are independent of the device thickness (confirmed for devices A, B, and C).

To further explore the temperature-dependent change in the magnetic exchange interactions we conducted fixed-frequency FMR experiments on a thin ($\sim 800\,$nm) lamella sample. 
In Fig.~\ref{fig4}e, we present FMR spectra recorded at 260 and 10\,K for two field directions each. The spectra were recorded while sweeping the field from the negative field-polarized state to the positive side and back, i.e., following the Hall-effect hysteresis curves. 
For further details, see also Supplementary Note\ 9. 
We also observe the hysteresis around 0.5\,T for $H \parallel c$ in the FMR data (see black curve in Fig.~\ref{fig4}e and inset). 
Furthermore, at 260\,K, we observe one narrow resonance mode attributed to the field-polarized state. As we tilt the field towards $H \parallel b$, the resonance preserves its narrow shape and the nearly isotropic angular dependence, varying between 0.62 and 0.75\,T. 
This suggests that above $T_\mathrm{SR}$, the magnetocrystalline anisotropy is weak. The maximum resonance field (corresponding to a magnetically hard axis) occurs around $24^\circ$. 
An approximate fit to the angle-dependent data yields a weak effective in-plane-anisotropy field, which is comparable in magnitude to the ASK compound Fe$_{1.9}$Ni$_{0.9}$Pd$_{0.2}$P \cite{Karube2021}. 
At temperatures well below $T_\mathrm{SR}$, the main resonance mode becomes broad for $H \parallel c$ and asymmetric. 
In Fig.~\ref{fig4}f, the error bars represent the resonance linewidth (half width at half maximum). 
Such a linewidth broadening is typical for the so-called field dragging, i.e., when $H$ is not parallel to $M$ due to a nonuniform magnetization for example. 
We observe that for high tilt angles close to the $b$ axis the resonances become narrower and shift to much lower fields of ~0.1 T. 
The overall angular dependence indicates that the anisotropy is much stronger below $T_\mathrm{SR}$. These significant changes in the magnetic anisotropies can explain the previously reported vanishing of ASKs and emergence of Bloch-type skyrmions at low temperatures \cite{Peng2020}. 
In addition, the observed hysteresis in transport at high fields indicates new hard-magnetic behavior for the in-plane field orientation. 
This, therefore, may provide the right environment of the establishment of other distinct skyrmionics, e.g., AFM bimerons \cite{Goebel2019}. 

\section*{CONCLUSION}
In summary, we demonstrated the direct detection of the Hall resistivity due to the emergence of ASKs in the Heusler compound Mn$_{1.4}$PtSn by combining electrical transport with magneto-optical microscopy, in line with our theoretical model. 
Our studies of transport, magnetization, and ferromagnetic resonance reveal a semi-hard magnetic phase, and, hence, the unique link of ASKs to anisotropies, not possible in the cubic B20 compounds with weak anisotropy. 
We find this semi-hard magnetic behavior to strengthen as we reduce the thickness of the sample, which is particularly interesting for scalable electronic devices \cite{He2020a,He2020b, Wu2020}. 
Many hard magnetic materials are strongly susceptible to finite sizes and may display nontrivial magnetic textures detectable by Hall transport \cite{Song2020, Sokolov2019}. 
As the thickness is decreased, the density of ASKs increases, which we observe directly using MFM, leading to an increase in the Hall-effect signature. 
Our theoretical prediction of the ASK THE magnitude matches the observed Hall signature values, thereby displaying the importance of the electronic structure for the unexpectedly large Hall signature. 
However, the disentanglement of the anomalous and topological Hall contributions remains a challenge, without spin-resolved visualization capabilities. 
In addition, the temperature dependence of the magnetic anisotropy, reflected in our Hall and FMR results, reveals the intimate link to the crystalline, electronic, and magnetic structure. 
Our atomistic model, solely based on the competing magnetic exchange interactions, captures the mechanism behind the emergence of ASKs in Mn$_{1.4}$PtSn. 
The novelty of tunability of texture sizes and the importance of the exchange interactions offers additional ways to manipulate and detect ASKs. 
Yet, smaller sizes and new materials with enhanced THE are desirable. 
Our observations offers a route to room-temperature skyrmionic applications based on multiple distinct types of skyrmionic textures.

\section*{METHODS}
{\bf Experimental Design.} The objectives of this study were to visualize the emergence of ASKs in mesoscale devices that allow a simultaneous investigation of the electrical transport under the influence of an external magnetic field and disentangle other contributions to the electrical-transport signature related to the complex magnetism of the multiskyrmion host compound Mn$_{1.4}$PtSn. 
We conducted polar magneto-optical Kerr microscopy as well as magnetic force microscopy on micron-sized samples fabricated by focussed-ion-beam assisted patterning. 
The former microscopy technique was successfully applied for the \textit{in-situ} detection of the Hall effect at room temperature. 
Furthermore, a detailed study of the ASK Hall signature depending on sample thickness, field orientation, and temperature was conducted. 
Chip-based FMR measurements depending on temperature, field, and angle were performed to obtain information on the ferromagnetic component of the complex magnetic background around temperatures at which the THE, associated with ASKs, is subsiding.

{\bf Crystal growth.} Single-crystals were prepared by flux-growth technique showing no microtwinning confirmed by Laue diffraction. 
Details can be found elsewhere \cite{Vir2019b}.

{\bf FIB microfabrication.} We fabricated transport devices from high-quality single crystals of Mn$_{1.4}$PtSn by the application of Ga or Xe FIB microstructuring, which enable high-resolution investigations of anisotropic transport. 
FIB micromachining has already proven extremely powerful in various other metallic compounds. 
A detailed description of the fabrication is provided in Supplementary Note\ 1 and 2. 
Further details on FIB-patterned transport devices can be found in previous works of some of the authors \cite{Moll2018,Ronning2017}. 
Thin (few microns thick) lamella-shaped slices of Mn$_{1.4}$PtSn were separated with FIB and manually transferred ex-situ onto a sapphire substrate into a thin layer of insulating two-component epoxy. 
We then deposited a 100\,nm thick gold (Au) layer on top in order to electrically connect the leads with the crystal. 
With the help of FIB, we thereafter patterned the Au interfaces into separate terminals. 
In a next step, we removed the Au from the central area of the slice before cutting the lamellae into Hall-bar-shaped transport devices, highlighted by purple color in Fig.~\ref{fig1}b. 
For device A, shown in Fig.~\ref{fig1}b, the thin slice of the crystal was polished stepwise to three different thicknesses by FIB before the transfer onto the substrate. 
This way, a microstructure device was created with three 3-point hall-bar devices connected in series.

{\bf Magnetotransport measurements.} We conducted electrical-transport measurements in a 14\,T Quantum Design physical properties measurement system by applying a standard a.c. lock-in technique. 
A SynkTek multichannel lock-in amplifier as well as Z\"urich Instruments lock-in amplifiers were used. The current is directed through the FIB cut microstructures (see purple colored parts in Fig.~\ref{fig1}b). 
Dimension of the devices were determined using a scanning electron microscope.

{\bf Magnetic force microscopy.} MFM measurements were performed in two instruments. 
For room-temperature measurements without external fields we used a Park Systems NX10 with MFM probes from Nanosensors at lift heights between 100 and 150\,nm. 
For measurements under magnetic fields, we used a AIST-NT SmartSPM 1000 with MFM probes from Nanosensors at lift heights between 250 and 350\,nm together with a Nd-based permanent magnet. 
The magnetic flux strength was measured at the position of the sample with a MAGSYS HGM09s magnetometer. 
All data analysis was performed with the Gwyddion software.

{\bf Polar magneto-optical Kerr microscopy.} Experiments were conducted with an AxioScope-type Carl Zeiss wide-field polarization microscope at room temperature. 
The sample was placed onto the pole of a solenoid-type, water-cooled electromagnet with a maximum magnetic field of 1.5\,T aligned along the $c$ axis. 
Special care was taken to remove parasitic Faraday contributions by the application of a motorized analyzer. 
Further details of the method can be found elsewhere \cite{Soldatov2018}.

{\bf Ferromagnetic Resonance.} A broadband frequency vector network analyzer (VNA) \cite{Goncalves2017} was used to probe the resonance properties of a Mn$_{1.4}$PtSn lamella $(20 \times 100 \times 0.8)\,\mu$m$^3$ prepared by FIB. 
A $20\,\mu$m wide coplanar wave guide was used for the transmission in an impedance-matched ($50\,\Omega$) chip-carrier design. 
The detection of the FMR modes was carried out via the change in the forward transmission parameter $S_{21}$ using a Keysight N5225A VNA. 
Field- and temperature-dependent measurements were performed in an Attocube DRY 1100 cryostat with a split-coil magnet at constant temperature and fixed frequency of 20\,GHz using 10\,dBm microwave input power. 
At each field step, the real and imaginary components of the S21 parameter were recorded, using 100 times averaging.

{\bf Numerical calculations.} First-principle calculations were performed on Mn$_{1.4}$PtSn by using the full-potential linearized augmented plane-wave code FLEUR following the work of Vir et al. \cite{Vir2019a}. 
From this, exchange parameters were extracted, where the largest interactions were $J_1$, $J_2$, $J_3$, and $D$. 
The atomistic spin calculations were performed in the code VAMPIRE for a lattice with $372 \times 372$ unit cells \cite{Swekis2021}. 
For computation efficiency the Heisenberg exchange parameters are decreased ($J_1 = 10\,$meV, $J_2 = 12.0\,$meV, $J_3 = -4\,$meV) and the DMI magnitude increased ($D_1 = 2.0\,$meV). 
The magnetocrystalline anisotropy is single axis and was chosen to be $0.5\,\mu$eV. 
To simulate the effect of the spin-reorientation transition, we set $J_3 = -2\,$meV. 
We used the Landau-Lifshitz-Gilbert equation to study the spin dynamics of the system.

\section*{ACKNOWLEDGEMENTS}
We would like to acknowledge K. Geishendorf from IFW for his help with transport-device fabrications.
Furthermore, we thank A. P. Mackenzie from MPI CPfS and R. Huebner from IBC, HZDR for their support.
We are thankful for the help of R. Narkovic and R. Illing from HZDR with the preparation of the substrate used in the FMR experiments.
We acknowledge the support of the HLD at HZDR, member of the European Magnetic Field Laboratory (EMFL).
We thank the department of A. Mackenzie at MPI CPfS in Dresden and the IBC at HZDR in Dresden for providing access and support to their FIB system.
B.E.Z.C. acknowledges support from the International Max Planck Research School for Chemistry and Physics of Quantum Materials (IMPRS-CPQM).
P.M. and L.M.E. acknowledges support by the German Research Foundation (DFG) under Grants No. EN 434/38-1 and MI 2004/3-1 as well as EN 434/40-1 as part of the Priority Program SPP 2137 “Skyrmionics”.
We acknowledge support by the Collaborative Research Center SFB 1143 (project-id 247310070) and the Würzburg-Dresden Cluster of Excellence on Complexity and Topology in Quantum Matter – \textit{ct.qmat} (EXC 2147, project-id 390858490).
P.J.W.M. acknowledges support by the European Research Council (ERC) under the European Union’s Horizon 2020 research and innovation program (grant no. 715730, MiTopMat).

\section*{AUTHOR CONTRIBUTIONS}
Single-crystal growth and characterization: PV\\
FIB microstructure fabrication: TH\\
Magnetotransport experiments: MW, SH, MU, RS, TH\\
SQUID VSM measurements: YH\\
MFM measurements: BEZC, PM\\
MOKE microscopy: IS, RS, YH\\
FMR measurements: FJTG, KL\\
First principle and atomistic spin dynamic simulations: JG\\
Supervision: TH, JW, CF, LME, RS, MK, KL, STBG, JG\\
Writing and editing of the original draft: All authors
 	
\section*{COMPETING INTERESTS}
The authors declare no competing interests.

\section*{DATA AVAILABILITY}
All data supporting the findings of this study are included in the main text and Supplementary Information. 
Raw data in ASCII format will be provided by the corresponding author upon reasonable request.

\section*{SUPPLEMENTARY MATERIALS}
In addition, we provide the following supplementary material:

Supplementary Notes (1 to 9)

Supplementary Figures (S1 to S16)

%Supplementary Movie 1

\section*{REFERENCES}
\bibliographystyle{naturemag}
\bibliography{Literature20211011}
\clearpage

\newpage
\newpage
\begin{figure}[tb]
	\centering
	\includegraphics[width=0.9\linewidth]{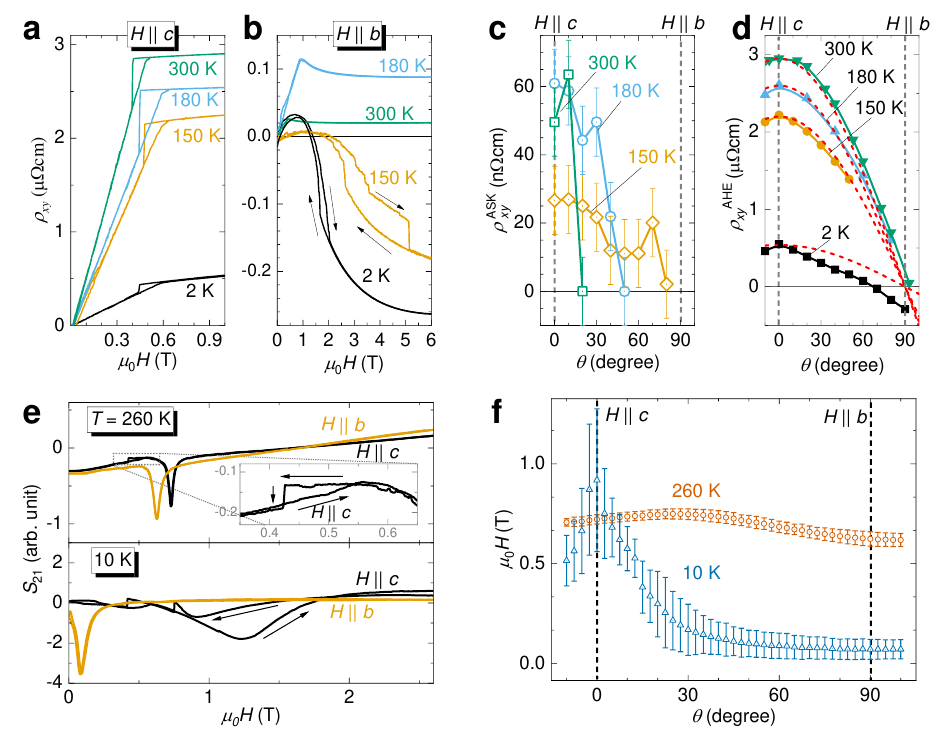}
	\caption{{\bf Angle-dependent Hall and FMR at various temperatures.} (a), (b) $\rho_{xy}$(H) of device B with $1\,\mu$m thickness recorded for $H \parallel c$ and $H \parallel a$, respectively, at various temperatures. (c) Extracted ASK Hall signature at three fixed temperatures for various angles. The error bars are defined as in Fig.~\ref{fig3}c. (d) Anomalous Hall coefficient extracted from linear fits to the high-field part above $5\,$T of $\rho_{xy}(H)$ plotted against the tilt angle $\theta$. Red dashed lines are fits with $\rho_{xy}(H)\propto \cos{\theta}$. (e), (f) FMR forward-transmission parameter, $S_{21}$. (e) FMR spectra for $H \parallel c$ and $H \parallel b$ recorded for a $0.8\,\mu$m thick lamella at 10 and 260\,K. (f) Angular dependence of the FMR field. Error bars represent the resonance linewidth values.
	}
	\label{fig4}
\end{figure}
\clearpage
\newpage
\newgeometry{left=0mm, right=0mm, top=0mm, bottom=0mm}  
\thispagestyle{empty}
%\begin{center}
\includegraphics[page=1, scale=1]{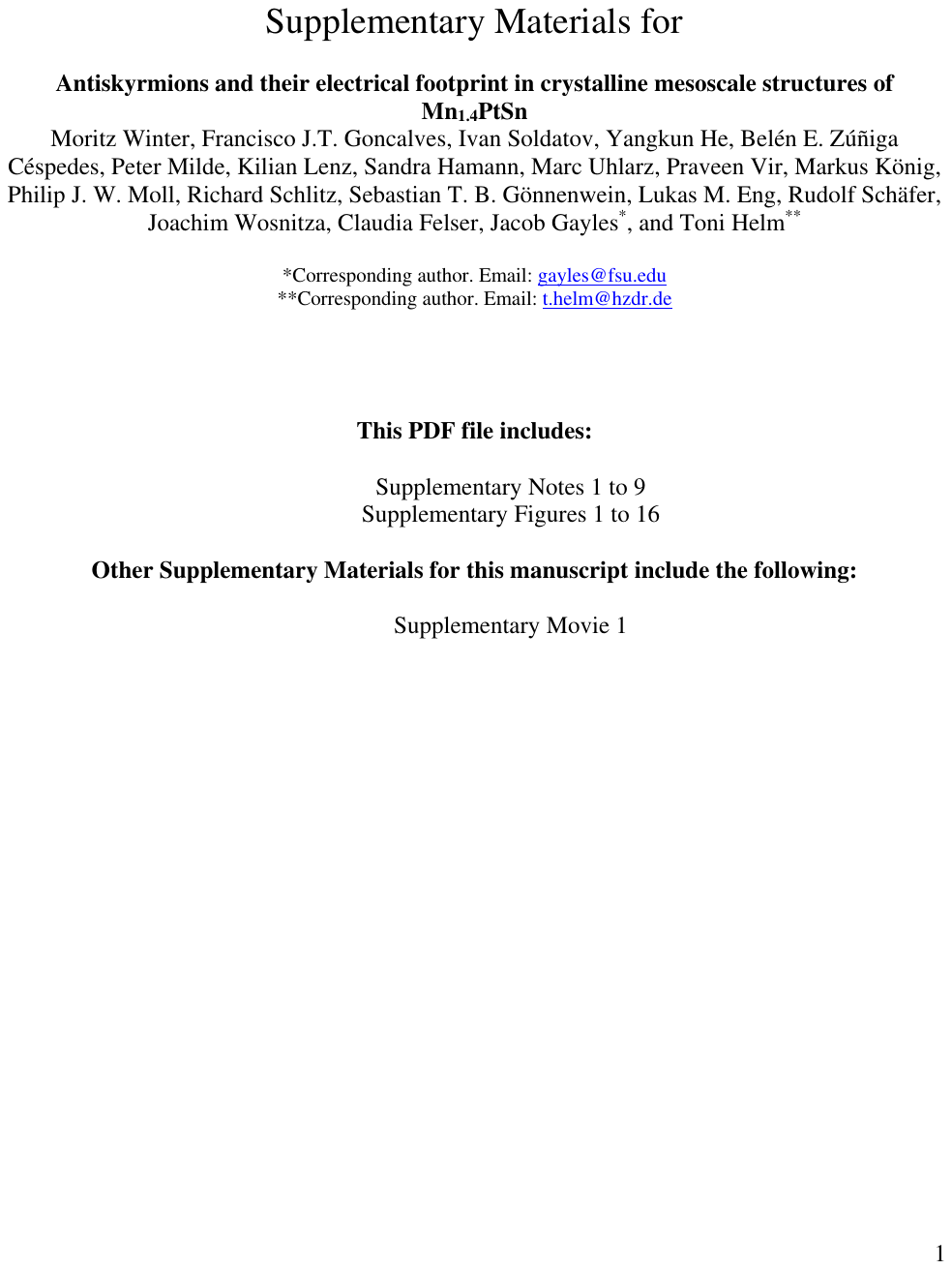}
%\end{center}
%\newpage
%\thispagestyle{empty} 
%\begin{figure}
	%\centering
	%\includegraphics[page=2, scale=1]{Winter_Supplement_Dec2022v1.pdf}
%\end{figure}
%\newpage
\includegraphics[page=3, scale=1]{Winter_Supplement_Dec2022v1.pdf}
\includegraphics[page=4, scale=1]{Winter_Supplement_Dec2022v1.pdf}
\includegraphics[page=5, scale=1]{Winter_Supplement_Dec2022v1.pdf}
\includegraphics[page=6, scale=1]{Winter_Supplement_Dec2022v1.pdf}
\includegraphics[page=7, scale=1]{Winter_Supplement_Dec2022v1.pdf}
\includegraphics[page=8, scale=1]{Winter_Supplement_Dec2022v1.pdf}
\includegraphics[page=9, scale=1]{Winter_Supplement_Dec2022v1.pdf}
\includegraphics[page=10, scale=1]{Winter_Supplement_Dec2022v1.pdf}
\includegraphics[page=11, scale=1]{Winter_Supplement_Dec2022v1.pdf}
\includegraphics[page=12, scale=1]{Winter_Supplement_Dec2022v1.pdf}
\includegraphics[page=13, scale=1]{Winter_Supplement_Dec2022v1.pdf}
\includegraphics[page=14, scale=1]{Winter_Supplement_Dec2022v1.pdf}
\includegraphics[page=15, scale=1]{Winter_Supplement_Dec2022v1.pdf}
\includegraphics[page=16, scale=1]{Winter_Supplement_Dec2022v1.pdf}
\includegraphics[page=17, scale=1]{Winter_Supplement_Dec2022v1.pdf}
\includegraphics[page=18, scale=1]{Winter_Supplement_Dec2022v1.pdf}
\includegraphics[page=19, scale=1]{Winter_Supplement_Dec2022v1.pdf}
\includegraphics[page=20, scale=1]{Winter_Supplement_Dec2022v1.pdf}
\includegraphics[page=21, scale=1]{Winter_Supplement_Dec2022v1.pdf}
\includegraphics[page=22, scale=1]{Winter_Supplement_Dec2022v1.pdf}
\end{document}